\documentclass[twocolumn,showpacs,preprintnumbers,amsmath,amssymb]{revtex4}
\usepackage{graphicx}
\usepackage{dcolumn}
\usepackage{bm}
\input{epsf}
\begin{document}
\title{{Anomalous transport: a deterministic approach.}}

\vskip4mm                      

\author{Roberto Artuso}
\email{Roberto.Artuso@uninsubria.it}
\altaffiliation{also at
Istituto Nazionale di Fisica Nucleare, Sezione di Milano,
Via Celoria 16, 20133 Milano, Italy} 
\author{Giampaolo Cristadoro}
\email{Giampaolo.Cristadoro@uninsubria.it}
\affiliation{Center for Nonlinear and Complex Systems and
Dipartimento di Scienze Chimiche, Fisiche e Matematiche, Universit\`a
dell'Insubria\\
Istituto Nazionale di Fisica della Materia, 
Unit\`a di Como, Via Valleggio 11, 22100 Como, Italy\\
22100 Como, Italy}
\date{\today}

\begin{abstract}

{We 
introduce a cycle-expansion (fully deterministic)
technique to compute the asymptotic behavior of
arbitrary order transport moments. The theory is applied to different 
kinds of one-dimensional
intermittent maps, and Lorentz gas with infinite horizon, confirming 
the typical appearance of phase transitions in the transport spectrum.

}
\end{abstract}

\maketitle
\narrowtext

Generic dynamical systems are characterized by the coexistence of chaotic 
regions and regular structures, and typical trajectories present 
regular segments, due to sticking to the ordered component of the phase 
space, separated by erratic behavior, due to wanderings in the chaotic 
sea. Though ubiquitous, this mixed behavior still involves hard 
theoretical problems, as present theories are tailored to fit
the two opposite paradigms of either integrable 
or fully chaotic systems. A particularly relevant feature associated to 
such weakly chaotic dynamics is anomalous transport:
important features of the process are captured by the function 
$\nu(q)$ that expresses the asymptotic growth of moments of arbitrary 
order:
\begin{equation}
\langle | x_t-x_0 |^q \rangle \, \sim \, t^{ \nu(q)} \, 
\label{nu-mom}
\end{equation}
ordinary diffusion yields  $\nu(q)=q/2$, while 
anomalous transport often leads to a non trivial behavior, which cannot 
be encoded by a single exponent, but rather typically exhibits a 
phase transition \cite{Ark,V-AD}.
Anomalous diffusion has been recognized as an ubiquitous phenomenon in recent years, from intermittency induced anomalous transport in 
one-dimensional maps \cite{Gei-AT} to the analysis of area-preserving 
maps in the presence of self-similar regular structures around 
accelerator modes \cite{Zas-AM}, from passive tracers dynamics in rotating flows \cite{Wee}, to charge carrier transport in amorphous semiconductors \cite{Mont}, to many others physically relevant contexts (see \cite{an-rev} and references therein).
In this paper we show how the transport exponents $\nu(q)$ may be 
computed in a crisply deterministic manner, by means of periodic orbit 
expansions \cite{DB,AACI}: we will then apply the formalism to various 
low dimensional settings, where analytic computations may be performed. 
These computations accomplish a twofold goal: besides checking the 
theory, they will also illustrate how subtle features of the underlying 
dynamical system are automatically included in the formalism.

For simplicity, we illustrate the analytic technique for the case of a one-dimensional  map on the real
line, even though the method is by no means limited to this context. The key property
of the dynamical system under investigation is represented by the symmetry properties
\begin{equation}
f(-x)  \, = \,   -f(x) \qquad \quad
f(x+n) \, = \, n+f(x)
\label{symm-P}
\end{equation}
for any $n \in \mathbf{N}$. These properties guarantee the absence of a net drift, as well as that
the map on the real line is obtained by lifting a circle
map $\hat{f}(\theta)$  (on the unit torus)
\begin{equation}
\hat{f}(\theta)\,=\, \left. f(\theta)\right|_{mod\,\,1} \qquad \theta \in {\mathcal{T}}=[0,1)
\label{tor-f}
\end{equation}
We may thus split the $f$ evolution into a box integer plus a fractional part: 
$x_n=N_n+\theta_n$,
where
\begin{equation}
N_{n+1} \,=\, N_{n} + \sigma(f(\theta_n)) \qquad \quad
\theta_{n+1}\, =\,   \hat{f}(\theta_n) \label{fracf}
\end{equation}
where $\sigma(y)=[y]$ is the denotes the integer part.  Transport properties
are accounted for by the generating function $G_n(\beta)$, defined as
\begin{equation}
\label{gen-f}
G_n(\beta)\,=\langle \,e^{\beta(x_n-x_0)} \rangle_0
\end{equation}
(the average being taken among initial conditions), whose asymptotic behavior may be
associated to the behavior of a generalized
\emph{dynamical zeta function}\cite{rD,DB,pD},
\begin{equation}
\label{gen-z}
G_n(\beta)\, \sim \, \frac{1}{2 \pi \imath}\int_{a - \imath \infty}^{a+ \imath \infty} \, ds\, e^{sn}
\, \frac{d}{ds} \ln \left[ \zeta_{(0)\beta}^{-1}(e^{-s}) \right]
\end{equation}
and $\zeta$ is expressed as an infinite product over prime periodic
orbits of the torus map $\hat{f}$:
\begin{equation}
\label{Dz}
\zeta_{(0)\beta}^{-1}(z)\,=\, \prod_{\{p\}}\,\left( 1 - \frac{e^{\beta \sigma_p}z^{n_p}}{|\Lambda_p|}
\right)
\end{equation}
The quantities that enter the definition (\ref{Dz}) are the prime period $n_p$ of the orbit $p$, its
instability $\Lambda_p=\prod_{i=0}^{n_p-1}\hat{f}'(\hat{f}^i(x_p)$ and the integer factor $\sigma_p$,
that accounts for the orbit's behavior once we unfold it on the real line. As a matter of fact, given
any point $x_p$ belonging to $p$ (that is $\hat{f}^{n_p}(x_p)=x_p$) we may either have that it is
a periodic point of the lift $f$ (i.e. $f^{n_p}(x_p)=x_p$), or it might be a running mode, 
$f^{n_p}(x_p)=x_p+\sigma_p$, with $\sigma_p \in \mathbf{Z}$ (see (\ref{tor-f})). If the
zeta function (\ref{Dz}) has a simple zero $z(\beta)$, then, from (\ref{gen-z}) it follows that 
the second moment of the distribution grows linearly in time (normal diffusion). This is generally the case with fully chaotic systems \cite{DB,rs,AACII,vb}: in the last
few years it has been realized \cite{gg,act,per,prisola} that weakly chaotic systems (in particular
one dimensional intermittent maps, or infinite horizon Lorentz gas models \cite{perL}) lead to more 
complicated analytic structure of the zeta function (\ref{Dz}), which typically exhibit branch 
points. In view of the inverse Laplace formula (\ref{gen-z}) the modified analytic structure
may induce anomalous behavior (nonlinear diffusion \cite{ACL,DB}).

\begin{figure}[t]
\centerline{\epsfxsize=4.cm \epsfbox{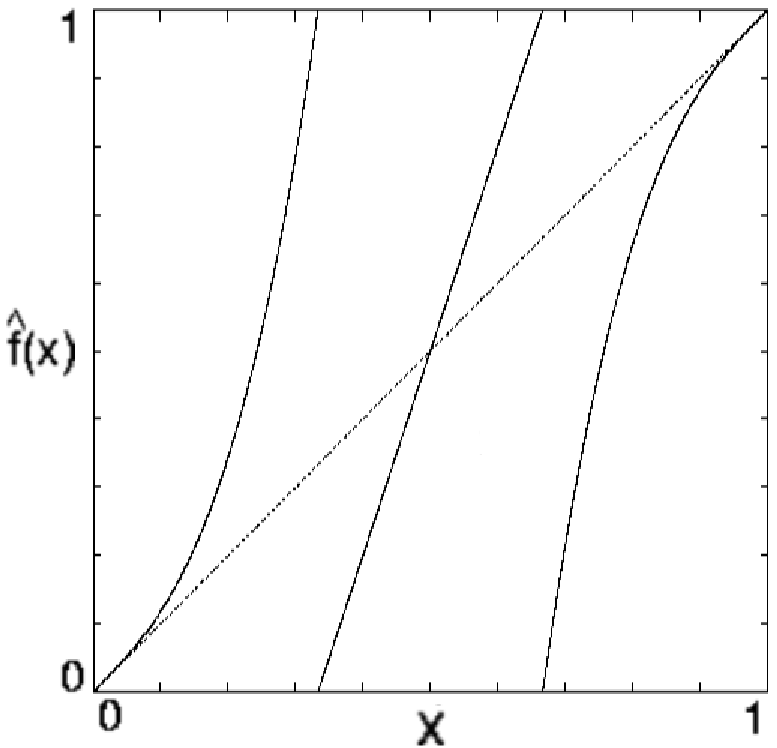}
\hspace{2mm}\epsfysize=4.cm \epsfbox{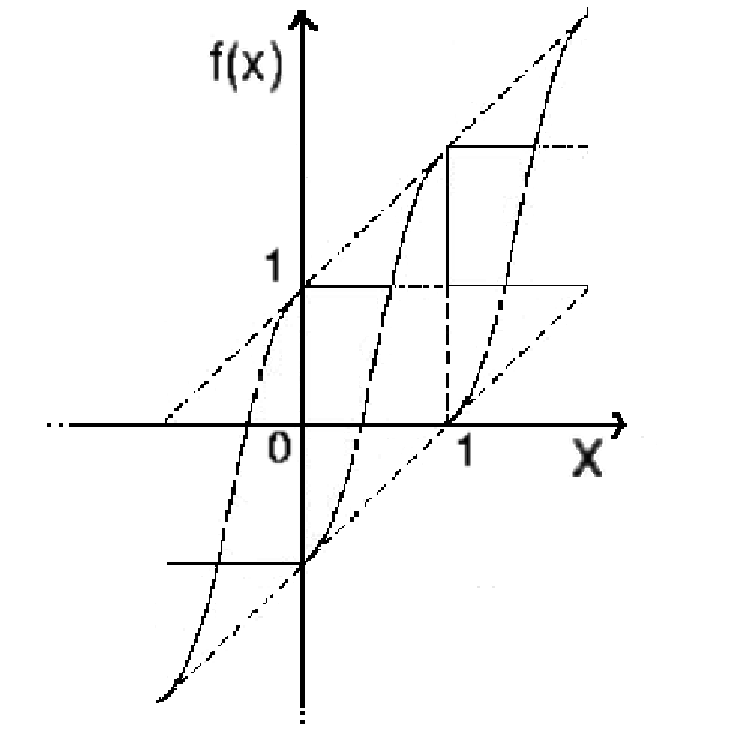}}
\vskip4mm
\caption{\label{f1}A torus map with intermittent fixed points, and the corresponding lift.}
\end{figure}
We address here the problem of going beyond diffusion, and characterize the whole spectrum 
of transport exponents $\nu(q)$ (\ref{nu-mom}). From small $\beta$ expansion of (\ref{gen-z}),
we get
\begin{eqnarray}
\sigma_k(n)&\,=\,& \langle (x_n -x_0 )^k \rangle_0 \,=\, \left. \frac{\partial^k \,} {\partial \beta^k}
G_n(\beta) \right|_{\beta=0} \nonumber \\
 &\,\sim\,& \left. \frac{\partial^k \,} {\partial \beta^k} \frac{1}{2 \pi \imath}\int_{a - \imath \infty}^{a+ \imath \infty} \, ds\, e^{sn}
\, \frac{d}{ds} \ln \left[ \zeta_{(0)\beta}^{-1}(e^{-s}) \right] \right|_{\beta=0} 
\label{gen-mom}
\end{eqnarray}
The evaluation of the integral on the right hand side of (\ref{gen-mom}) requires dealing with
high order derivatives of a composite function: this is accomplished by making use of Fa\`a di Bruno formula:
\begin{eqnarray} 
\frac{d^n \,}{dt^n}H(L(t))= \sum_{k=1}^n\,\sum_{k_1,\cdots k_n}
\, \frac{d^k H}{dt^k}  (L(t))\cdot B_{\vec{k}}(L(t))
\label{fdb}\\
B_{\vec{k}}(L(t))=\left(\frac{1}{1!}\frac{dL}{dt}\right)^{k_1} \cdots \left(\frac{1}{n!}\frac{d^nL}{dt^n}\right)^{k_n} \, \, \, \,  \quad \, ~\\
\vec{k}=\{k_1,\dots k_n\} \,\text{with}\, \sum k_i=k ,\quad \sum i\cdot k_i=n 
\end{eqnarray}

When the analytic structure of the zeta function is known, from (\ref{fdb})
we may single out the leading singularity in the logarithmic derivative, and then
estimate the asymptotic behavior of (\ref{gen-mom}), for instance by employing Tauberian
theorems for Laplace transforms \cite{Fell}. 

We will apply the technique to two classes of one dimensional maps, where deviations from 
fully chaotic behavior are provided by marginal fixed points, of intermittent type \cite{PM}.
The torus map is shown in fig. (\ref{f1}): it consists of three branches, the central one being hyperbolic
(with constant slope), while the other two include a marginal fixed point: to unfold it on the real
line it is sufficient to assign jumping numbers to branches. For instance we get the lift of fig. (\ref{f1})
once we assign $\sigma_{-}=-1$ to the left branch, $\sigma_{c}=0$ to the central one and 
$\sigma_{+}=+1$ to the right branch. 
The map is actually
taken as a straightforward generalization of the Gaspard-Wang piecewise linear approximation
\cite{plapp} of Pomeau-Manneville map: the intermittent behavior is determined by dynamics near
the pair of parabolic fixed points, which is accounted for by the intermittency exponent $\gamma > 1$,
(the map goes like $x_{n+1}\sim x_n + x_n^{\gamma}$ near the origin, with an analogous behavior
at the twin fixed point in $x=1$).
The 
zeta function can then be written as
\begin{equation}
\label{pl-dz}
\zeta_{(0)\beta}^{-1}(z)\,=\, 1-az-bz\sum_{k=1}^{\infty}\, \frac{z^k}{k^{\alpha+1}}\cosh (\beta k)
\end{equation}
where $a$ and $b$ are fixed by specifying the central region slope and the normalization
condition $\zeta_{(0)0}^{-1}(1)=0$, and 
\begin{equation}
\label{stab-z}
\alpha=1/(\gamma-1).
\end{equation} 
The appearance of the Bose function
$g_{\mu}(z)\,=\,\sum_{l=1}^{\infty}\, \frac{z^l}{l^{\mu}}$
is due to sequences of orbits coming closer and closer to the marginal fixed points: their 
stability increases only polynomially with the period \cite{plapp,AACII}, a clear signature
of local deviation from typical hyperbolic behavior (which is ruled by exponential instability
growth). To estimate the various 
contributions in (\ref{fdb}) we 
remind the 
behavior as $z \to 1^{-}$
\begin{eqnarray}
g_{\mu}(z) \sim  \left\{
\begin{array}{ll}
(1-z)^{\mu -1} \quad & \mu<1 \\
\ln (1-z) \quad & \mu=1 \\
\zeta(\mu)+C_{\mu}(1-z)^{\mu-1}+D_{\mu}(1-z) \quad & \mu \in (1,2) \\
\zeta(2)+C_2(1-z) \ln (1-z) \quad &  \mu=2 \\
\zeta(\mu)+C_{\mu}(1-z) \quad & \mu> 2
\end{array} \right.
\label{B-as}
\end{eqnarray}
and moreover take into account that
\begin{eqnarray}
\left. \frac{\partial^i \, }{\partial \beta^i}\zeta_{(0)\beta}^{-1}(z)\right|_{\beta=0}\,\sim \,
\left\{
\begin{array}{ll}
0 \qquad & i \,\,\mathrm{odd} \\
zg_{\alpha+1-i}(z) \quad & i\,\, \mathrm{even}
\end{array} 
\right.
\label{djz}
\end{eqnarray}
Now take a generic term in (\ref{fdb}): and denote it by ${\cal D}_{k_1 \dots k_n}$: we have
in view of (\ref{djz})
\begin{equation}
\label{typ-fdb}
{\cal D}_{k_1 \dots k_n}\,\sim\, \frac{1}{(\zeta_{(0)0}^{-1}(z))^k}\prod_j\,(g_{\alpha+1-j}(z))^{k_j}
\,=\,\frac{{\cal D}^{+}_{k_1 \dots k_n}}{{\cal D}^{-}_{k_1 \dots k_n}}
\end{equation}
where the ${\cal D}^{+}$ picks up the contributions from the product of Bose functions, and
all $j$ must be even, due to (\ref{djz}).
First we consider the case $\alpha \in (0,1)$, which corresponds to 
$\gamma > 2$: we have ${\cal D}^{-}_{k_1 \dots k_n}\sim (1-z)^{k\alpha}$, that,
together with (\ref{B-as}), implies that the dominant singularity is of the form
\begin{equation}
\label{ddom}
{{\cal D}_n}\,\sim \, \frac{1}{(1-z)^{\rho}}
\end{equation}
where $\rho$ is determined by 
\begin{equation}
\label{rho}
\rho=  \sup_{\{k_1 \dots k_n\}} ( k\alpha + \sum_j (j-\alpha)k_j) =n
\end{equation} 
Once plugged into (\ref{gen-mom}) this leads to the estimate
$\label{nu-norm}
\nu(q)\,=\, q
$,
which means that the whole set of moments is ruled by ballistic behavior (at least
for even exponents, where the method applies).
We now turn to the more subtle case $ \alpha > 1$: since the dynamical zeta function has a
simple zero we get ${\cal D}^{-}_{k_1 \dots k_n}\sim (1-z)^k$, while the terms appearing in
${\cal D}^{+}$ modify the singular behavior near $z=1$ only for sufficiently high $j$
\begin{equation}
g_{\alpha+1-j}(z)\,\sim\,
\left\{
\begin{array}{ll}
(1-z)^{\alpha -j}\quad & j> \alpha \\
\zeta(\alpha +1-j) \quad & j< \alpha
\end{array}
\right.
\label{ggg}
\end{equation}
If all $\{j\}$ are less than $\alpha$ then the singularity is determined by ${\cal D}^{-}$: keeping in
mind that the highest $k$ value is achieved by choosing $j=2$ and $k_2=n/2$, we get, by 
proceeding as before
\begin{equation}
\label{mom-}
\nu(q)\,=\,\frac{q}{2} \qquad q < \alpha
\end{equation}
When $q$ exceeds $\alpha$ we have to take into account possible additional singularities in
${\cal D}^{+}$, and thus we get
\begin{equation}
\label{ddom2}
{{\cal D}_n}\,\sim \, \frac{1}{(1-z)^{\rho}}
\end{equation}
where $\rho$ is determined by 
\begin{equation}
\rho= \sup_{\{k_1 \dots k_n\}} ( k + \sum_{j>\alpha} (j-\alpha)k_j) =
\left\{
\begin{array}{ll}
n/2 \quad & n < 2(\alpha-1) \\
n+1-\alpha \quad & n> 2(\alpha-1)
\end{array}
\right.
\label{rho2}
\end{equation} 
which, once we take (\ref{mom-}) into account, yields
\begin{equation}
\label{momfin}
\nu(q)\,=\,
\left\{
\begin{array}{ll}
q/2 \quad & q< 2(\alpha-1) \\
q+1-\alpha \quad & q> 2(\alpha-1)
\end{array}
\right.
\end{equation}
The set of exponents thus has a nontrivial structure, characterized by a sort of phase transition
for $q=2(\alpha-1)$, a rather universal feature of many systems exhibiting anomalous transport
\cite{V-AD}. We notice that the parameter ruling the presence of a phase transition (and the
explicit form of the spectrum) is $\alpha$, the exponent describing the polynomial 
instability growth of periodic orbits coming closer and closer to the marginal
fixed point, and thus describing the sticking to the regular part of the phase space: in the present
example $\alpha=\/(\gamma-1)^{-1}$, and thus the sticking exponent is easily connected to the
intermittency index.

\begin{figure}
\centerline{\epsfxsize=5.cm \epsfbox{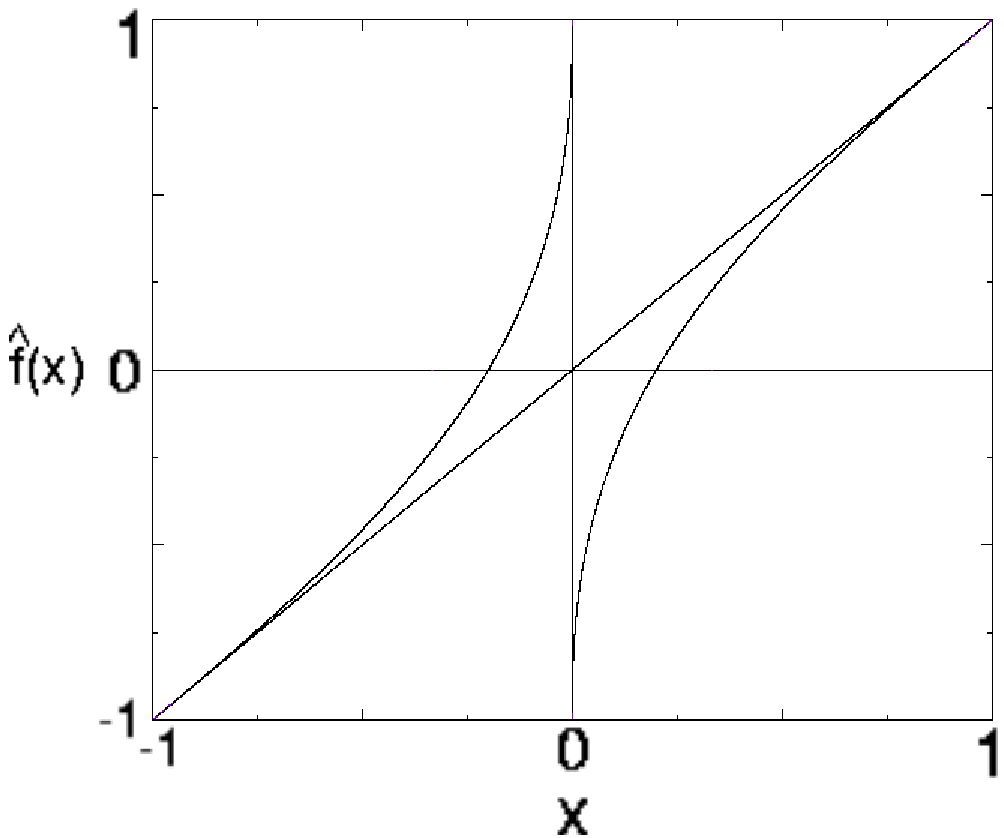}}
\vskip4mm
\caption{\label{f3}The intermittent map with constant invariant measure.}
\end{figure}
We now discuss a further 1-d example: first we recall that the former torus map has non trivial ergodic properties: an absolutely
continuous invariant measure only exists for $\alpha > 1$ (see for instance \cite{ci} and references 
therein): the ergodic behavior is much more complex when $\alpha <1$. 
In the new example
\cite{Ark}  while the functional form of the map near marginal
fixed points is identical to the former case, ergodic properties are completely different (the measure is not singular). 
The torus map, again dependent on an intermittency parameter $\gamma$,
 is implicitly defined on $\mathcal{T}=[-1,1)$ in the following way \cite{Ark}:
\begin{equation}
x\,=\,
\left\{
\begin{array}{ll}
\frac{1}{2\gamma}\left(1+ \hat{f}(x)\right)^{\gamma} \qquad & 0 < x < 1/(2\gamma) \\
\hat{f}(x) + \frac{1}{2\gamma} \left( 1- \hat{f}(x) \right)^{\gamma} \qquad & 1/(2\gamma) < x < 1
\end{array}
\right.
\label{arkmap}
\end{equation}
for negative values of $x$ the map is defined as $\hat{f}(-x)=-\hat{f}(x)$ (cfr (\ref{symm-P})):
see fig. (\ref{f3}).
\begin{figure}[t]
\centerline{\epsfxsize=5.cm \epsfbox{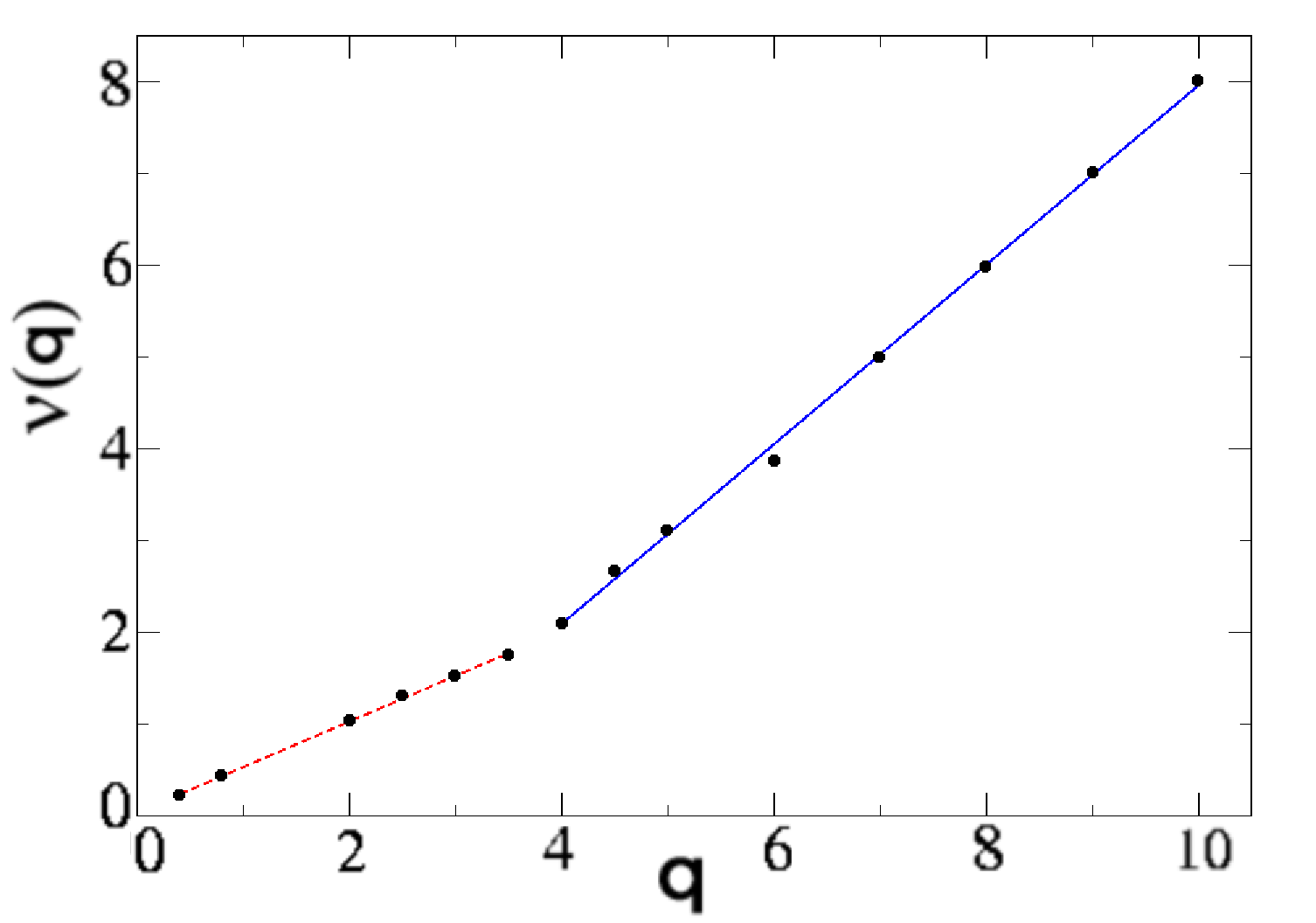}}
\vskip4mm
\caption{\label{f4}Spectrum of the transport moments for the map (\ref{arkmap}) with $\gamma=1.5$: 
the best fit on numerical data is $y=0.50 x +0.04$ for the dotted line, and $y=0.98x+1.82$ for
the full line.}
\end{figure}
The peculiar ergodic features of the map, namely the existence of a {\em constant} invariant
measure for {\em any} value of $\gamma$ arise from the property
$
\sum_{y=\hat{f}^{-1}(x)}\, 1/{\hat{f}'(y)}\,=\,1
$
Also in this case we can easily identify families of orbits coming closer and closer to the marginal
fixed points, but evaluating their instability requires some care, as the slope in the chaotic region
is not bounded from above. A piecewise linear approximation (in the same spirit as \cite{plapp})
is still possible \cite{acprep}, but we must put particular attention on matching the summation property
:
the corresponding dynamical zeta function is
\begin{equation}
\zeta_{(0)\beta}^{-1}(z)\,=\,1-\zeta(\alpha+2)z\sum_{k=1}^{\infty}\, \frac{z^k}{k^{\alpha+2}}
\cosh(\beta k)
\label{zark}
\end{equation}
where again $\alpha=1/(\gamma-1)$. Thus the continuity of the measure deeply modifies the relationship
between the intermittency exponent and the instabilities of periodic orbits shadowing the 
marginal fixed points: by repeating the steps of our former calculation we get that we always get a phase transition with
\begin{equation}
\label{arkspec}
\nu(q)\,=\,
\left\{
\begin{array}{ll}
q/2 \quad & q< 2\alpha \\
q-\alpha \quad & q> 2\alpha
\end{array}
\right.
\end{equation}
which may also be checked numerically (see fig. (\ref{f4})).

Our last example shows how the proposed technique may be applied to higher dimensional systems, where a detailed layout of the full symbolic dynamics is beyond our present understanding. We consider the Lorentz gas with infinite horizon (square lattice of circular scatterers where a test particle freely moves, colliding elastically with the disks). The mechanism that in this case leads to deviation from normal behavior is due to the possibility of arbitrarily long flights of free (collisionless) motion along corridors. In this example the reduced dynamics is that of a Sinai billiard: while a complete description of the full set of periodic orbits of such a system is not known yet a a family of periodic orbits approaching closer and closer the infinite free flights is singled out: it consists (for the unfolded billiard) to orbits that jump $n$ lattice spacings between collisions: their number and instability can be evaluated by geometric arguments\cite{perL}, yielding an istability that grows polynomially with $n$ with a power $3$: so their role is analogous to modes leading to Bose functions in (\ref{pl-dz}) with $\alpha=2$. In view of (\ref{momfin}) this leads to $\nu(q)=q/2$ for $q< 2$ and $\nu(q)=q-1$ for $q > 2$, like recently suggested in \cite{L-Ott}: a careful analysis, based on (\ref{B-as}) moreover gives $\sigma_2(n) \sim n \ln n$ (see \cite{Ble}), so that our method is also capable of picking up logarithmic corrections to the dominating power-law behavior.

We have proposed a crisply deterministic technique to investigate the full spectrum of
transport exponent for chaotic systems: in particular this method is capable of explaining
the different phase transitions that possibly arise, without any direct information on the invariant
measure: the only information to be plugged is local growth of instabilities near the 
marginal structures, together with the appropriate jumping factor.

This work was partially supported by INFM PA project {\em
Weak chaos: theory and applications}, and by EU contract QTRANS Network 
({\em Quantum transport on an atomic scale}).


\end{document}